\documentclass[amssymb,amsfonts,amsmath,reprint,aps,prapplied,floatfix]{revtex4-1}
\usepackage{graphicx}
\usepackage{hyperref}

\newcommand{\mrm}[1]{\mathrm{#1}}

\newcommand{\abs}[1]{\left| #1 \right|}

\begin{document}

\title{Fast and Ultrasensitive Electrometer Operating at the Single-Photon Level}
\date{\today}

\author{B.~L.~Brock}
\email{Benjamin.L.Brock.GR@dartmouth.edu}
\altaffiliation[\\Present address: ]{Department of Applied Physics, Yale University, New Haven, Connecticut 06520, USA}
\affiliation{Department of Physics and Astronomy, Dartmouth College, Hanover, New Hampshire 03755, USA}

\author{Juliang Li}
\altaffiliation[Present address: ]{High Energy Physics Divison, Argonne National Laboratory, 9700 South Cass Avenue, Argonne, IL 60439, USA}
\affiliation{Department of Physics and Astronomy, Dartmouth College, Hanover, New Hampshire 03755, USA}

\author{S.~Kanhirathingal}
\affiliation{Department of Physics and Astronomy, Dartmouth College, Hanover, New Hampshire 03755, USA}

\author{B.~Thyagarajan}
\affiliation{Department of Physics and Astronomy, Dartmouth College, Hanover, New Hampshire 03755, USA}

\author{M.~P.~Blencowe}
\affiliation{Department of Physics and Astronomy, Dartmouth College, Hanover, New Hampshire 03755, USA}

\author{A.~J.~Rimberg}
\email{Alexander.J.Rimberg@dartmouth.edu}
\affiliation{Department of Physics and Astronomy, Dartmouth College, Hanover, New Hampshire 03755, USA}

\begin{abstract}
We demonstrate fast and ultrasensitive charge detection with a cavity-embedded Cooper pair transistor (cCPT) via dispersive readout of its Josephson inductance.  We report a minimum charge sensitivity of $14$~$\mu e/\sqrt{\mathrm{Hz}}$ with a detection bandwidth on the order of $1$ MHz using $16$ attowatts of power, corresponding to the single-photon level of the cavity.  In addition, our measured sensitivities are within a factor of $5$ of the quantum limit for this device.  The single-photon-level sensitivity of the cCPT is comparable to that of the rf-SET, which typically operates using picowatts of power corresponding to hundreds of thousands of photons in its tank circuit.  Our results support the feasibility of using the cCPT to mediate an optomechanical interaction that reaches the single-photon strong coupling regime.
\end{abstract}

\maketitle


Fast and ultrasensitive electrometers have been instrumental to the advancement of basic science.  They have been used to detect in real time the tunneling of electrons in a quantum dot \cite{Lu2003}, determine the tunneling rates of quasiparticles in superconducting devices \cite{Naaman2006}, and search for signatures of Majorana zero modes in nanowires \cite{Zanten2020}.  In addition, the rapid detection of single electrons is crucial for the readout of quantum-dot-based qubits \cite{Petta2005}, for which operating at lower photon numbers reduces measurement backaction \cite{D'Anjou2019}.  In this same vein, ultrasensitive electrometers are at the heart of many schemes for sensing the displacement of charged mechanical resonators \cite{Knobel2003, LaHaye2004, Naik2006}, as well as for coherently coupling mechanical resonators to microwave cavities \cite{Rimberg2014, Heikkila2014, Pirkkalainen2015}.  To observe and take advantage of quantum effects in such hybrid systems it is often essential that their coupling be strong at the single-photon level, a regime that has been achieved for quantum dots \cite{Mi2016, Mi2018} but not yet for mechanical resonators despite significant effort \cite{Zoepfl2020, Schmidt2020, Kounalakis2020, Bera2021}.  Reaching the single-photon strong optomechanical coupling regime, where a single cavity photon causes sufficient radiation pressure to displace the mechanical resonator by more than its zero-point uncertainty, would enable the generation of nonclassical states of both light and motion \cite{Nunnenkamp2011, Rabl2011}, as well as provide a rich platform for studying the quantum-to-classical transition and other fundamental physics \cite{Aspelmeyer2014}.

Electrometers based on the single electron transistor (SET) are among the fastest and most sensitive reported in the literature to date.  Radio-frequency single electron transistors (rf-SETs) are the best known of these devices, having achieved sensitivities below $1$~$\mu e/\sqrt{\mrm{Hz}}$ \cite{Brenning2006} and bandwidths greater than $100$ MHz \cite{Schoelkopf1998}.  The rf-SET encodes the charge gating the SET in the power dissipated by the SET, which is embedded in a tank circuit to enable RF readout of this dissipation.  This dissipative detection typically requires picowatts of power, corresponding to hundreds of thousands of photons in the tank circuit, rendering the rf-SET unsuitable for some of the aforementioned applications and making it impossible to integrate the rf-SET with modern near-quantum-limited amplifiers \cite{Castellanos-Beltran2007, Macklin2015, Sivak2019} (which typically saturate well below the picowatt scale).  Dispersive electrometers based on the SET have also been developed, which encode the gate charge in the resonant frequency of a tank circuit.  Such electrometers have been operated using femtowatts of power \cite{Sillanpaa2004,Naaman2006}, corresponding to tens or hundreds of photons, and have achieved sensitivities as low as $30$~$\mu e/\sqrt{\mrm{Hz}}$ \cite{Sillanpaa2005}.  More recently, dispersive gate-based sensors have been developed \cite{Gonzalez-Zalba2015} that have surpassed the performance of SET-based electrometers.  These devices have achieved sensitivities as low as $0.25$~$\mu e/\sqrt{\mrm{Hz}}$ with bandwidths approaching $1$ MHz using $100$ attowatts of power, corresponding to hundreds of photons \cite{Schaal2020}.  

In this letter we demonstrate ultrasensitive dispersive charge detection with a cavity-embedded Cooper pair transistor (cCPT) \cite{Brock2021_characterization,Kanhirathingal2021}.  Using $16$ attowatts of power, corresponding to the single-photon level of the cavity, we measure a minimum charge sensitivity of $14$~$\mu e/\sqrt{\mathrm{Hz}}$.  We find that the cCPT operates within a factor of $5$ of its theoretical quantum-limited sensitivity, this discrepancy being due to frequency noise, amplifier noise, and the nonlinearity of the device.  Another limitation of the present device is quasiparticle poisoning \cite{Aumentado2004}, which prevents us from studying the cCPT at its theoretically-optimal operating point.  Based on these results we expect an optimized sample could achieve a sensitivity as low as $0.4$~$\mu e/\sqrt{\mathrm{Hz}}$, rivaling that of the best gate-based sensor \cite{Schaal2020}.  Due to its ability to operate at the single-photon-level, the cCPT has been proposed as a platform for reaching the single-photon strong coupling regime of optomechanics \cite{Rimberg2014}.  Our results support the feasibility of this proposal and represent an important step toward its realization.

Here we study the same device characterized experimentally in Ref. \cite{Brock2021_characterization}.  The most important parameters of this realization of the cCPT are shown in Table \ref{tab:ccpt_parameters}.  For more information on this device, including sample images, fabrication methods, and characterization techniques, see Ref. \cite{Brock2021_characterization}.

\begin{figure}[!t]
\includegraphics{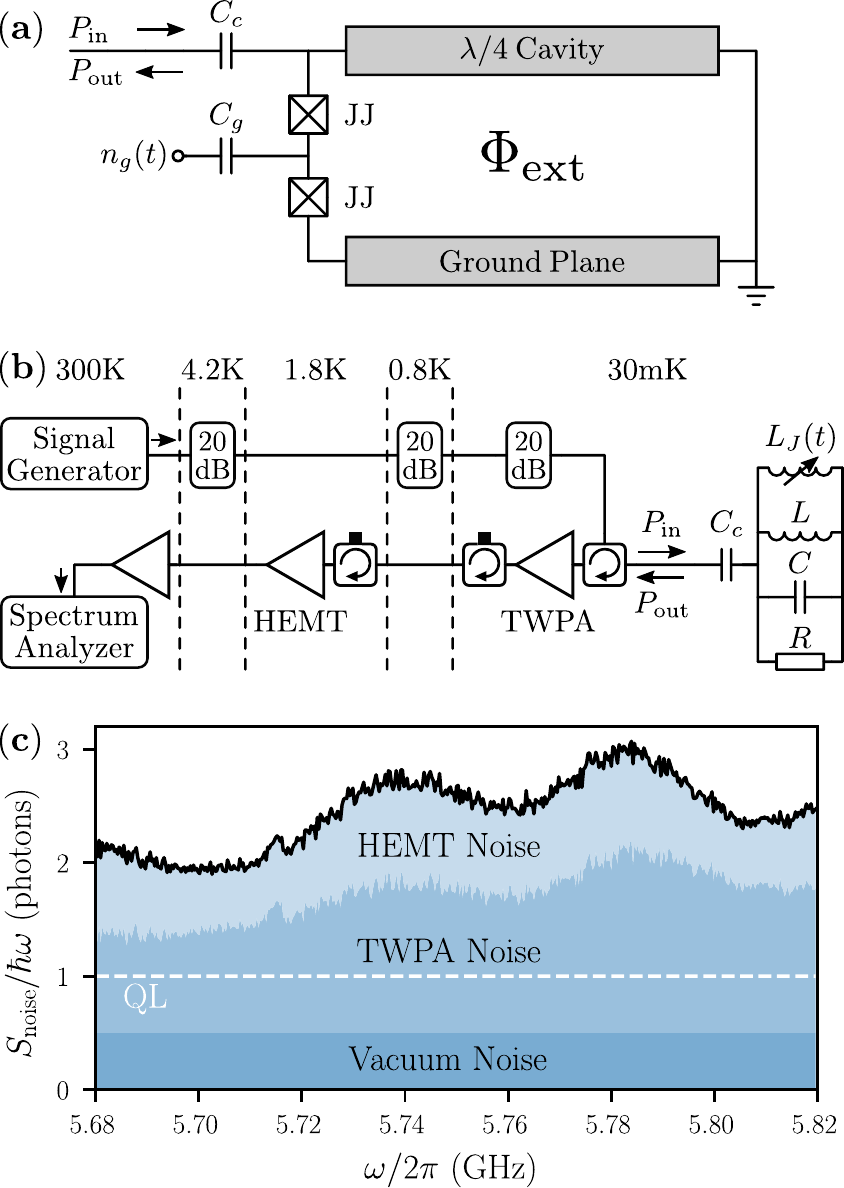}
  \caption{(a) Schematic of the cCPT. (b) Schematic of the measurement circuitry.  The cavity behaves as a parallel RLC circuit when driven near its fundamental frequency \cite{text_pozar}, and the CPT behaves as an inductance $L_{J}$ in parallel with the cavity.  (c) System noise referred to the sample plane (solid black line).  Shaded areas show the contribution of each noise source.  The dashed white line is the quantum limit.}
  \label{fig:schematics_and_noise}
\end{figure}

The cCPT, depicted schematically in Fig. \ref{fig:schematics_and_noise}(a), has two components: a quarter-wavelength ($\lambda/4$) coplanar waveguide cavity and a Cooper pair transistor (CPT).  The CPT consists of two Josephson junctions (JJs) with an island between them that can be gated via the capacitance $C_{g}$.  The CPT is connected between the voltage antinode of the cavity and the ground plane, such that the two form a SQUID loop.  Embedded in this way, the CPT behaves as a nonlinear Josephson inductance $L_{J}$ in parallel with the cavity that can be tuned by both the number of electrons $n_{g}$ gating the island and the flux $\Phi_{\mathrm{ext}}$ threading the SQUID loop.  The gate charge $n_{g}$ is thus encoded in the resonant frequency $\omega_{0}$ of the cavity, which can then be detected via microwave reflectometry.  The theoretical charge sensitivity of the cCPT in this mode of operation is derived from first principles in Ref. \cite{Kanhirathingal2021}.  This device can be operated at much lower powers than comparable SET-based dispersive electrometers \cite{Naaman2006, Sillanpaa2004, Sillanpaa2005} for two key reasons.  First, we use a distributed superconducting microwave cavity rather than a lumped-element LC circuit, yielding much lower dissipation.  Second, we can tune the CPT band structure via the external flux $\Phi_{\mrm{ext}}$, which provides us greater flexibility in biasing the device to an optimally-sensitive point.

\begin{table}[t]
  \begin{tabular}{|c|c|} \hline
  Josephson energy & $E_{J}/h = 14.8$ GHz  \\ \hline 
  Charging energy  & $E_{C}/h = 54.1$ GHz \\ \hline
  Gate capacitance & $C_{g} = 6.3$ aF \\ \hline
  Coupling capacitance & $C_{c} = 7.1$ fF \\ \hline
  Bare cavity frequency & $\omega_{\lambda/4}/2\pi \approx 5.76$ GHz \\ \hline
  Cavity linewidth & $\kappa_{\mrm{tot}}/2\pi \approx 1.4$ MHz \\ \hline
  Cavity length & $\ell = 5135 \mu$m \\ \hline
  Characteristic impedance & $Z_{0} = 50$ $\Omega$ \\ \hline
  \end{tabular}
\caption{Parameters of the cCPT \cite{Brock2021_characterization}.}
\label{tab:ccpt_parameters}
\end{table}


To measure the charge sensitivity of the cCPT we drive the cavity with a resonant carrier signal while modulating the gate about a dc bias point $n_{g}$ such that $n_{g}(t) = n_{g} + \sqrt{2}(q_{\mrm{rms}}/e)\cos(\omega_{g}t)$, which in turn modulates the resonant frequency according to $\omega_{0}(t) = \omega_{0} + \sqrt{2}(\partial\omega_{0}/\partial n_{g})(q_{\mathrm{rms}}/e)\cos(\omega_{g}t)$.  As a result, the reflected carrier signal is phase-modulated leading to output power $P_{\mrm{out}}$ proportional to $q_{\mrm{rms}}^{2}$ at the sideband frequencies $\omega_{0}\pm\omega_{g}$.  Thus, given the rms charge modulation amplitude $q_{\mrm{rms}}$, we can use a spectrum analyzer with resolution bandwidth $B$ to measure the sidebands and thereby extract the charge sensitivity $\delta q$ from
\begin{equation}\label{eq:empirical_sensitivity}
\delta q = \frac{q_{\mrm{rms}}}{\sqrt{2B}\times 10^{\mrm{SNR}/20}},
\end{equation}
where $\mrm{SNR}$ is the single-sideband signal to noise ratio expressed in decibels \cite{Brenning2006, Aassime2001}.  Here we consider the total power at the two sidebands to be the signal of interest, since it is possible to combine them via homodyne mixing, leading to the factor of $1/\sqrt{2}$ above.  Theoretically, the output sideband power $P_{\mrm{out}}(\omega_{0}\pm\omega_{g})$ can be expressed
\begin{equation}\label{eq:output_sideband_power}
P_{\mrm{out}}(\omega_{0}\pm\omega_{g}) = \frac{2\kappa_{\mrm{ext}}^{2}}{\kappa_{\mrm{tot}}^{2}(\omega_{g}^{2} + \kappa_{\mrm{tot}}^{2}/4)}\abs{\frac{q_{\mrm{rms}}}{e}\frac{\partial\omega_{0}}{\partial n_{g}}}^{2}P_{\mrm{in}},
\end{equation}
where $\kappa_{\mrm{ext}}$ and $\kappa_{\mrm{tot}}$ are the external and total damping rates of the cavity, respectively, and $P_{\mathrm{in}}$ is the input carrier power at the plane of the sample \cite{Brock2021_characterization}.  The theoretical charge sensitivity can therefore be expressed as \cite{Kanhirathingal2021}
\begin{equation}\label{eq:theoretical_charge_sensitivity}
\delta q = \frac{\kappa_{\mrm{tot}}}{2\kappa_{\mrm{ext}}}\sqrt{\frac{S_{\mrm{noise}}}{P_{\mrm{in}}}\left(\omega_{g}^{2} + \frac{\kappa_{\mrm{tot}}^{2}}{4}\right)}\abs{\frac{\partial\omega_{0}}{\partial n_{g}}}^{-1}e.
\end{equation}
To evaluate this expression we use the sample-referred $S_{\mrm{noise}}$ and $P_{\mrm{in}}$ (discussed below), as well as the values of $\kappa_{\mrm{ext}}$, $\kappa_{\mrm{tot}}$, and $\omega_{0}(n_{g}, \Phi_{\mrm{ext}})$ determined from a detailed characterization of the device \cite{Brock2021_characterization}.  The damping rates are approximately $\kappa_{\mrm{ext}}/2\pi\approx 1.2$ MHz and $\kappa_{\mrm{tot}}/2\pi\approx 1.4$ MHz, though these depend on $\omega_{0}(n_{g},\Phi_{\mrm{ext}})$ and can vary by $10\%-20\%$.  The corresponding quantum-limited sensitivity of the device is obtained by evaluating Eq. \eqref{eq:theoretical_charge_sensitivity} at the quantum limit of system noise for our measurement scheme, $S_{\mrm{noise}}^{\mrm{QL}}=\hbar\omega$, as discussed below.

Importantly, both Eqs. \eqref{eq:empirical_sensitivity} and \eqref{eq:theoretical_charge_sensitivity} are only valid when $q_{\mrm{rms}}/e\ll \omega_{g}/(\partial\omega_{0}/\partial n_{g})$, which ensures that the amplitude of the resulting frequency modulation is small compared to $\omega_{g}$ and that $P_{\mrm{out}}(\omega_{0}\pm\omega_{g}) \propto q_{\mrm{rms}}^{2}$.  In all of our measurements we use sufficiently small $q_{\mrm{rms}}$ to satisfy this constraint.  Furthermore, Eq. \eqref{eq:theoretical_charge_sensitivity} is most accurate in the linear response regime for which $n\ll\kappa_{\mrm{tot}}/|K|$, where $n=4\kappa_{\mrm{ext}}P_{\mrm{in}}/\hbar\omega_{0}\kappa_{\mrm{tot}}^{2}$ is the average number of intracavity photons and $K$ is the Kerr nonlinearity of the cCPT \cite{Brock2021_characterization}.  Experimentally, we find that for $n\ll\kappa_{\mrm{tot}}/|K|$ the output sideband power grows linearly with $P_{\mrm{in}}$ as expected from Eq. \eqref{eq:output_sideband_power}, but as $n$ approaches $\kappa_{\mrm{tot}}/|K|$ this trend becomes sub-linear.  Near this threshold, $P_{\mrm{out}}(\omega_{0}\pm\omega_{g})$ begins to decrease with increasing $P_{\mrm{in}}$.  For the present device this threshold corresponds to the single-photon-level \cite{Brock2021_characterization}, so we perform all of our measurements with $n\lesssim 1$.  In this sense the single-photon-level operation of the cCPT can be viewed as both an enabling feature (for the reasons described earlier) and a constraint, but this constraint could be avoided in future devices by changing $E_{J}$, $E_{C}$, and $\kappa_{\mrm{tot}}$.

The detection bandwidth of the present device, which determines the maximum rate at which the cavity can respond to changes in $n_{g}$, is set by $\kappa_{\mrm{tot}}$ and is on the order of $1$ MHz.  The bandwidth can be improved by increasing the coupling capacitance $C_{c}$, thereby increasing $\kappa_{\mrm{ext}}$, but this also affects the single-photon-level charge sensitivity.  Setting $n=1$ and assuming negligible internal loss such that $\kappa_{\mrm{tot}}\approx \kappa_{\mrm{ext}}$, Eq. \eqref{eq:theoretical_charge_sensitivity} predicts $\delta q \propto \sqrt{\kappa_{\mrm{tot}}}$ for $\omega_{g}\ll\kappa_{\mrm{tot}}$.  However, if we restrict ourselves to the linear-response regime rather than the single-photon level we can operate with $n \sim \kappa_{\mrm{tot}}/\abs{K}$, in which case $\delta q$ is independent of $\kappa_{\mrm{tot}}$.  Lastly, it is worth noting that if one increases $\kappa_{\mrm{ext}}$, one also increases the charge noise coupling to the cCPT via the input-output transmission line \cite{Kanhirathingal2021}, which is negligible in the present device.

The cCPT is housed in a dilution refrigerator with a base temperature of $T\lesssim 30$ mK and measured using the circuitry depicted schematically in Fig. \ref{fig:schematics_and_noise}(b), which is nearly identical to that used in Ref. \cite{Brock2021_characterization}.  The one difference here is that we use a near quantum-limited TWPA \cite{Macklin2015} as a first-stage amplifier.  We use the techniques described in Ref. \cite{Brock2021_characterization} to refer all input and output powers, as well as the system noise $S_{\mrm{noise}}(\omega)$, to the plane of the sample.  The measured system noise, shown in Fig. \ref{fig:schematics_and_noise}(c), is due to the half-photon of vacuum noise $S_{\mrm{vac}}=\hbar\omega/2$ in the input/output transmission line \cite{Clerk2010} and the added noise of our amplifier chain $S_{\mrm{amp}}$, such that $S_{\mrm{noise}} = S_{\mrm{vac}}+S_{\mrm{amp}}$.  For all of the charge sensitivity measurements we report, the noise floor near the sideband frequencies is dominated by this system noise, which is why we use the same notation for these two quantities.  At sufficiently low gate modulation frequencies, however, the noise floor is dominated by $1/f$ charge noise \cite{Paladino2014}.  This regime occurs below about $1$ kHz in our case \cite{Brock2021_characterization}.  We determine the noise added by the TWPA and HEMT independently by measuring the gain of the amplifier chain and total system noise twice: once with the TWPA pump on and once with it off.  Over the operating range of the cCPT (between $5.68$ GHz and $5.82$ GHz), the TWPA contributes $1.2$ photons of noise ($50\%$ of total) while the HEMT contributes $0.7$ photons ($30\%$ of total) on average.  The room temperature amplifier contributes negligibly to the sample-referred system noise $S_{\mrm{noise}}$.  The quantum limit of noise in this system is one photon, such that $S_{\mrm{noise}}^{\mrm{QL}} = \hbar\omega$, since phase-insensitive amplifiers must add at least a half-photon of noise \cite{Caves1982}.  Thus, our average system noise is only a factor of $2.4$ greater than the quantum limit for this measurement scheme, such that the theoretical sensitivity (Eq. \ref{eq:theoretical_charge_sensitivity}) is only a factor of $\sqrt{2.4}$ greater than the quantum-limited sensitivity.  


\begin{figure}[!t]
\includegraphics{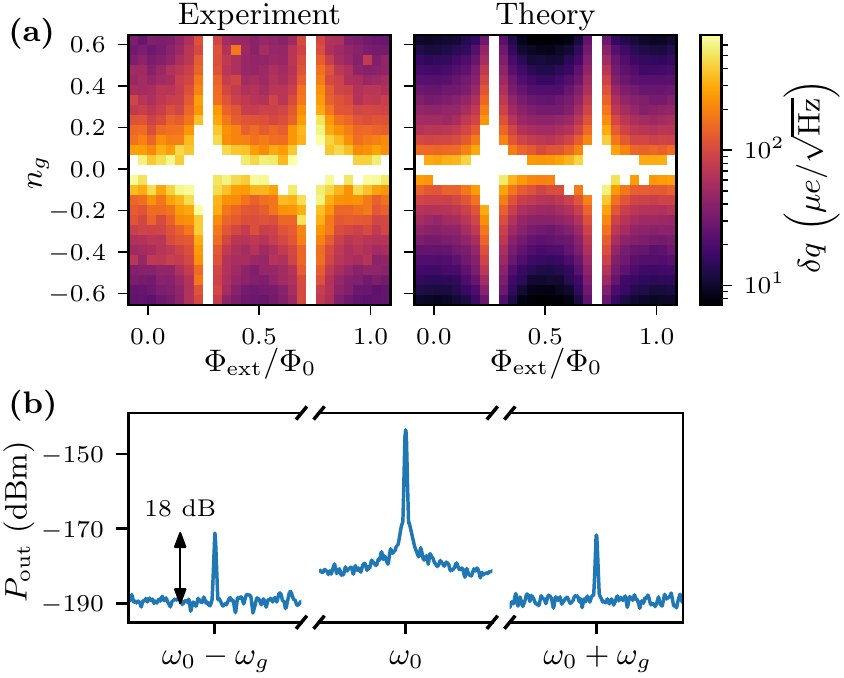}
  \caption{(a) Measured and theoretical charge sensitivities, obtained using Eqs. \eqref{eq:empirical_sensitivity} and \eqref{eq:theoretical_charge_sensitivity} respectively, as a function of gate and flux.  Data is omitted where the sidebands could not be resolved from the noise floor. (b) Sample-referred spectrum analyzer trace of the optimal charge sensitivity measurement, corresponding to $\delta q = 14$~$\mu e/\sqrt{\mrm{Hz}}$.  The carrier frequency is $\omega_{0}/2\pi = 5.806$ GHz, the gate modulation frequency is $\omega_{g}/2\pi = 350$ kHz, the span of each segment is $1$ kHz, and the resolution bandwidth is $B=10$ Hz.  The noise floor near the carrier is due to $1/f$ charge noise \cite{Brock2021_characterization,Paladino2014}.}
  \label{fig:combined_charge_sensitivity}
\end{figure}

In order to compare the cCPT's charge sensitivity with its theoretical performance, given by Eq. \eqref{eq:theoretical_charge_sensitivity}, we first measure $\delta q$ as a function of both the gate charge $n_{g}$ and external flux $\Phi_{\mrm{ext}}$.  Although we can access a full period of $\Phi_{\mrm{ext}}$ (from $0$ to the magnetic flux quantum $\Phi_{0}$), we can only access the gate range $-0.65<n_{g}<0.65$ due to quasiparticle poisoning \cite{Brock2021_characterization}.  We perform this measurement using an input power $P_{\mrm{in}}= -141$ dBm $\approx 8$ aW and gate modulation amplitude $q_{\mrm{rms}} = 10^{-3}e$.  Ideally we would set $\omega_{g}$ to be significantly less than $\kappa_{\mrm{tot}}/2\approx 2\pi\times 700$ kHz to minimize Eq. \eqref{eq:theoretical_charge_sensitivity}, but in our experiments we observe cross-talk between our gate and flux lines at frequencies below about $650$ kHz.  We therefore use $\omega_{g}/2\pi = 800$ kHz, such that the gate modulation does not also induce a flux modulation.  To measure the reflected power and noise floor at $\omega_{0}\pm\omega_{g}$ we use a resolution bandwidth $B=1$ Hz.  

The results of this measurement are shown in Fig. \ref{fig:combined_charge_sensitivity}(a).  We find that the variation of $\delta q$ with $n_{g}$ and $\Phi_{\mrm{ext}}$ is in good agreement with theory, but our measured sensitivities are about $3$ times worse than theory.  We attribute this discrepancy to two factors.  First and foremost, the resonant frequency fluctuates due to $1/f$ charge and flux noise \cite{Brock2021_characterization,Brock2020} over the course of each measurement, which means our carrier is not always on resonance.  On average, this reduces the output sideband power yielding worse charge sensitivity than expected.  Second, we used a sufficiently high input power that $P_{\mrm{out}}(\omega_{0}\pm\omega_{g})$ scales sublinearly with $P_{\mrm{in}}$ due to the Kerr nonlinearity.  Although this improves the sensitivity overall and was necessary to resolve the sidebands over a large area of the gate/flux parameter space, it causes the measured sensitivity to diverge from theory since the latter assumes $P_{\mrm{in}} \propto P_{\mrm{out}}(\omega_{0}\pm\omega_{g})$.  Finally, since $S_{\mrm{noise}}/S_{\mrm{noise}}^{\mrm{QL}} \approx 2.4$, the factor of $3$ discrepancy between theory and experiment means our measured sensitivities are within a factor of $5$ of the quantum limit.  In this measurement we find a minimum charge sensitivity of $24$~$\mu e/\sqrt{\mrm{Hz}}$ at $(n_{g},\Phi_{\mrm{ext}}) = (0.63, 0.0)$, whereas our predicted theoretical and quantum-limited sensitivities at this point are $9$~$\mu e/\sqrt{\mrm{Hz}}$ and $6$~$\mu e/\sqrt{\mrm{Hz}}$, respectively.

In order to optimize $\delta q$ we narrow our search to the gate range $0.6\leq |n_{g}|\leq 0.65$ and the flux points $\Phi_{\mrm{ext}}=0,\Phi_{0}/2$.  At these flux points the resonant frequency of the cCPT is insensitive to flux, so we can reduce our gate modulation frequency to $\omega_{g}/2\pi = 350$ kHz without the gate/flux cross-talk interfering with our results.  To maintain a small frequency modulation amplitude relative to $\omega_{g}$, we also reduce $q_{\mrm{rms}}$ to $5\times 10^{-4}e$.  For this measurement we use a resolution bandwidth $B=10$ Hz.  

\begin{table}[t]
  \begin{tabular}{|c|c|c|c|} \hline
  \textbf{Electrometer} & \begin{tabular}{@{}c@{}}$\delta q$ ($\mu e/\sqrt{\mrm{Hz}}$) \end{tabular} & \begin{tabular}{@{}c@{}} $P_{\mrm{in}}$ (aW) \end{tabular} & \begin{tabular}{@{}c@{}} $n$ photons \end{tabular}  \\ \hline 
 cCPT* & $14$ & $16$ & $1$  \\ \hline
 Best gate sensor\cite{Schaal2020}* & $0.25$ & $100$ & $190$ \\ \hline
 Best rf-SET\cite{Brenning2006} & $0.9$ & $6\times 10^{6}$ & $2\times 10^{5}$ \\ \hline
 Andresen et al.\cite{Andresen2008} & $2.3$ & $3\times 10^{8}$ & $2\times 10^{6}$ \\ \hline
 L-SET\cite{Sillanpaa2005}* & $30$ & $1\times 10^{4}$ & $70$  \\ \hline
 Naaman et al.\cite{Naaman2006}* & $52$ & $2\times 10^{3}$ & $150$  \\ \hline
 Bell et al.\cite{Bell2012}*  & $70$ & $3\times 10^{7}$ & $2\times 10^{5}$  \\ \hline
 rf-QPC\cite{Cassidy2007} & $200$ & $1\times 10^{9}$ & $7\times 10^{7}$ \\ \hline
  \end{tabular}
\caption{Comparison of the cCPT with a representative set of fast and ultrasensitive electrometers.  Asterisks indicate dispersive electrometers.}
\label{tab:comparing_electrometers}
\end{table}

We find a minimum charge sensitivity of $14$~$\mu e/\sqrt{\mrm{Hz}}$ at $(n_{g},\Phi_{\mrm{ext}}) = (0.625,0.0)$ using an input power $P_{\mrm{in}} = -138$ dBm $\approx 16$ aW.  Under these conditions our predicted theoretical and quantum-limited sensitivities are $5$~$\mu e/\sqrt{\mrm{Hz}}$ and $3$~$\mu e/\sqrt{\mrm{Hz}}$, respectively.  The spectrum analyzer trace of this optimal measurement is shown in Fig. \ref{fig:combined_charge_sensitivity}(b).  At this bias point the resonant frequency is $\omega_{0}/2\pi = 5.806$~GHz, the external damping is $\kappa_{\mrm{ext}}/2\pi = 1.24$~MHz, and the total damping is $\kappa_{\mrm{tot}}/2\pi = 1.62$~MHz, such that the number of intracavity photons is $n = 4\kappa_{\mrm{ext}}P_{\mrm{in}}/\hbar\omega_{0}\kappa_{\mrm{tot}}^{2} \approx 1$.  This single-photon-level sensitivity is rivaled only by gate-based sensors \cite{Schaal2020}, rf-SETs \cite{Brenning2006}, and carbon nanotube-based rf-SETs \cite{Andresen2008}, all of which operate with orders of magnitude more photons.  In Table \ref{tab:comparing_electrometers} we compare the performance of the cCPT to a representative set of fast (detection bandwidth $\gtrsim 1$ MHz) and ultrasensitive ($\delta q < 10^{-3} e/\sqrt{\mrm{Hz}}$) electrometers.  Clearly, the cCPT is unparalleled in its ability to operate at low powers and photon numbers.    As discussed earlier, this makes it ideal for mediating an optomechanical interaction that reaches the single-photon strong coupling regime \cite{Rimberg2014}.


There remains significant room for improving the sensitivity of the cCPT, with two distinct approaches for doing so.  The most promising approach is to reduce quasiparticle poisoning (QP) \cite{Aumentado2004}, which prevents us from operating at gate biases above $|n_{g}|\approx 0.65$ \cite{Brock2021_characterization}.  If we were able to operate the present device at $(n_{g},\Phi_{\mrm{ext}}) = (0.9,\Phi_{0}/2)$ we would expect to attain a charge sensitivity of $\delta q \approx 0.4$~$\mu e/\sqrt{\mrm{Hz}}$, assuming the same factor of $3$ discrepancy with theory as we observe experimentally.  The present device was designed with a $9$ nm thick CPT island \cite{Brock2021_characterization} to suppress QP \cite{Yamamoto2006}, but other fabrication techniques could be employed to reduce it further.  These include oxygen-doping the CPT island \cite{Aumentado2004} and embedding quasiparticle traps near the CPT \cite{Rajauria2012}.  The other approach is to mitigate the discrepancy between our measured sensitivities and the quantum limit.  One such improvement would be to use a truly quantum-limited amplifier, which would improve our sensitivities by a factor of $\sqrt{S_{\mrm{noise}}/\hbar\omega}~\approx~\sqrt{2.4}$.  Another such improvement would be to stabilize the resonant frequency against $1/f$ noise using a Pound-locking loop \cite{Lindstroem2011}.  It may also be possible to improve the sensitivity of the cCPT by exploiting the nonlinearity of the cCPT \cite{Laflamme2011,Tosi2019} or incorporating a parametric drive near $2\omega_{0}$ \cite{Krantz2016}.

Several important applications exist for single-photon-level charge sensing with the cCPT.  First and foremost, the cCPT can be used to dispersively sense any quantity that can be tied to electrical charge, two notable examples being the spin state of quantum-dot-based qubits and the position of a charged nanomechanical resonator.  For quantum-dot-based qubits, the spin states can be encoded in charge states via spin-to-charge conversion \cite{Petta2005,Hanson2007}.  For a charged nanomechanical resonator, the position of the resonator can be encoded in the charge on a capacitor \cite{Knobel2003,LaHaye2004,Naik2006}.  In both cases, the measurement backaction on the relevant degree of freedom is proportional to the number of photons in the cavity, such that single-photon-level operation is preferable \cite{D'Anjou2019,Clerk2010}.  Second, the cCPT can be readily integrated with near-quantum-limited amplifiers \cite{Castellanos-Beltran2007, Macklin2015, Sivak2019}, which typically saturate well below the level of power required by rf-SETs \cite{Brenning2006,Schoelkopf1998} and rf-QPCs \cite{Cassidy2007}.  Finally, the cCPT has been proposed as a platform for mediating an optomechanical interaction that reaches the single-photon strong coupling regime \cite{Rimberg2014}.  Our demonstration of single-photon-level electrometry with the cCPT supports the feasibility of this proposal and represents an important step toward its realization.

\begin{acknowledgments}
We thank W. F. Braasch for helpful discussions and W. Oliver for providing the TWPA used in these measurements.  The sample was fabricated at Dartmouth College and the Harvard Center for Nanoscale Systems.  B.L.B., S.K., and A.J.R. were supported by the National Science Foundation under Grant No. DMR-1807785.  J.L. was supported by the Army Research Office under Grant No. W911NF-13-1-0377.  M.P.B. was supported by the National Science Foundation under Grant No. DMR-1507383.  
\end{acknowledgments}

\end{document}